\documentclass{emulateapj}

\usepackage{color}
\usepackage{times}
\usepackage{fancyhdr}
\usepackage{float}
\usepackage{alltt}
\usepackage{listings}
\usepackage{enumerate}
\usepackage[T1]{fontenc}
\usepackage{ae}
\usepackage{rotating}
\usepackage{subfigure}
\usepackage{amsmath}
\usepackage{amssymb}
\usepackage{graphicx}
\usepackage{dcolumn}
\usepackage{bm}
\usepackage{aas_macros_pk}
\usepackage{natbib}

\def\be{\begin{equation}}
\def\ee{\end{equation}}
\def\beg{\begin{align}}
\def\eeg{\end{align}}
\def\bi{\begin{itemize}}
\def\ei{\end{itemize}}
\def\ben{\begin{enumerate}[1.]}
\def\een{\end{enumerate}}

\def\deg{\ensuremath{^{\circ}}}

\newcommand{\bo}{\raise-1mm\hbox{\Large$\Box$}}

\newcommand{\solarmass}{M_{\odot}c^2}

\newcommand\given{\,|\,}
\newcommand\signal{\mathrm{signal}}
\newcommand\noise{\mathrm{noise}}
\newcommand\diff{\, \mathrm{d}}

\keywords{Gravitational waves, Ultraviolet: general, X-rays: general}

\shorttitle{Swift follow-up observations of candidate gravitational-wave transient events}
\shortauthors{The LIGO Scientific and Virgo Collaborations et al.}

\begin{document}

\title{Swift follow-up observations of candidate gravitational-wave transient events}

\author{
P.~A.~Evans\altaffilmark{1},
J.~K.~Fridriksson\altaffilmark{2},
N.~Gehrels\altaffilmark{3},
J.~Homan\altaffilmark{2},
J.~P.~Osborne\altaffilmark{1},
M.~Siegel\altaffilmark{4},
A.~Beardmore\altaffilmark{1},
P.~Handbauer\altaffilmark{5},
J.~Gelbord\altaffilmark{4},
J.~A.~Kennea\altaffilmark{4}
% ...
}

% ...
\altaffiltext{1}{Department of Physics and Astronomy, University of Leicester, Leicester, LE1 7RH, UK}
\altaffiltext{2}{MIT Kavli Institute for Astrophysics and Space Research,
77 Massachusetts Avenue, Cambridge, MA 02139, USA}
\altaffiltext{3}{NASA Goddard Space Flight Center, Greenbelt, MD 20771, USA}
\altaffiltext{4}{The Pennsylvania State University, University Park, PA  16802, USA }
\altaffiltext{5}{E\"otv\"os Lor\'and University, Budapest, 1117 Hungary}

% ...

\begin{abstract}
We present the first multi-wavelength follow-up observations of two candidate 
gravitational-wave (GW) transient events recorded by LIGO and Virgo
in their 2009-2010 science run. The events were selected with low latency
by the network of GW detectors (within less than 10 minutes) and their
candidate sky locations were observed by the Swift observatory
(within 12 hours). Image transient detection
was used to analyze the collected electromagnetic data, which
were found to be consistent with background. Off-line analysis of the
GW data alone has also established that the selected 
GW events show no evidence of an astrophysical origin;
one of them is consistent with background and the other one was a
test, part of a ``blind injection challenge''.
With this work we demonstrate the feasibility of rapid follow-ups of
GW
transients and establish the sensitivity improvement joint electromagnetic
and GW observations could bring.
This is a first step toward an electromagnetic follow-up program in
the regime of routine detections with the advanced GW
instruments expected within this decade. In that regime multi-wavelength
observations
will play a significant role in completing the astrophysical identification
of GW sources.
We present the methods and results from
this first combined analysis and discuss its implications in terms of 
sensitivity for the present and future instruments. 
\end{abstract}

\pacs{ }

\maketitle

\color{white}{.}
\color{black}
\clearpage

\section{Introduction}

\setcounter{footnote}{0}
Some of the key questions in the pursuit of sources of transient gravitational
radiation detectable by LIGO \citep{LIGOinstrument} and
Virgo \citep{Virgoinstrument} relate to their electromagnetic (EM)
signatures and our ability to observe them \citep{s6emmethods}.  In several of these sources, like core collapse supernovae and
neutron star-neutron star or neutron star-black hole mergers,
energetics suggest gravitational waves (GWs) are likely
to be accompanied by EM emission across the spectrum and over time
scales ranging from seconds to days
\citep{Fryer2002, Piran2005, Meszaros2006, Nakar2007, Corsi2009}.
Multi-wavelength EM  observations of such events have already set the
paradigm for improved constraints on source astrophysics set jointly
rather than separately. 
Prompt outbursts, as well as afterglows in X-ray, optical and radio
associated with Gamma-Ray Bursts (GRBs) and supernovae have shed light
on the progenitors and the astrophysics of these systems
\citep{Kulkarni1998,Bloom1999,Matheson2003,Berger2005,Gehrels2005,Soder2005,BergerRev2009,Zhang2009,Soder2010,Berger2011}.
The benefit of coupling them to GW observations will be tremendous as 
this will bring the GW observations into astrophysical and cosmological
context.
Besides increasing detection confidence \citep{kochanek}, multi-wavelength observations may
improve source localization down to the arcsecond level, leading to
identification of the host galaxy and measurement of the redshift
\citep{Schutz1986,Sylvestre2003,Stubbs2008,Phinney2009,Stamatikos2009,Bloometal2009,Metzger2010,Metzger2012}.
Conversely, the absence of any EM signature for an otherwise confident
transient detection with the GW detectors alone will provide
constraints on emission mechanisms, progenitors and energetics.

LIGO \citep{LIGOinstrument},
and Virgo \citep{Virgoinstrument} form a network of
interferometric detectors aiming to make the first direct observations
of GWs. In their 2009-2010 data-taking period
this network consisted of three interferometers:
LIGO-Hanford in
Washington State in the USA, LIGO-Livingston in Louisiana in the USA,
and Virgo in Italy.
During these science runs of the instruments we implemented a first program
that could allow prompt EM follow-up of candidate GW transients.
Starting with GW data, low-latency searches for
compact binary star coalescences \citep{CBClowlatency} and un-modeled 
GW transients were performed.
This allowed the prompt identification and sky localization of GW
candidates, which for the first time were passed on to ground-based
telescopes and Swift in order to be followed up.
In an earlier publication \citep{s6emmethods} we presented the details
of the implementation 
and testing of this low latency search and its coupling to
EM astronomy.
In this paper we report on the analysis of the Swift data we collected as part
of this program.

With the start of the 2009-2010 LIGO-Virgo runs we established a
Target-of-Opportunity (ToO) program with Swift in order to search for possible
afterglow in X-ray,
ultraviolet and optical wavelengths of a small number of GW
transient candidates. The sensitivity of the GW detectors at the time made the
chance of a detection small, but non-negligible. 
For the case of binary sources with at least one
neutron star in the system, the rate of detectable merger events was predicted 
to be in the  range of 2.7$\times$10$^{-4}$ 
to 0.3 events per year \citep{cbcrates}. This ToO program
was exercised twice;  neither time led to detection of an EM 
counterpart to a GW transient.
Nonetheless, the program addressed implementation questions and established the first
joint observation and coordinated data analysis by the LIGO-Virgo network and
Swift satellite. 

In this paper we present the results from the EM follow-up program
involving only Swift ---
results from the sister program involving the follow-up of GW candidates by
ground-based optical and radio telescopes will be the subject of a
forthcoming publication. 
The paper is organized as follows. In Section 2 we describe
the Swift observatory. 
We then review in Section 3 the procedure for targeting EM follow-up
of GW events, as described in detail in \cite{s6emmethods}.
The Swift observations and analysis of data are described in Section 4.
In Section 5 we present our formalism for combining results from the joint
GW-EM search, including the simulation work we performed.
We conclude in Section 6 with a discussion and outlook for this kind
of joint search.

\section{The Swift observatory} 

The Swift Gamma-Ray Burst Mission \citep{Gehrels}, developed and launched under NASA's Medium Explorer Program,
is unique for its broad wavelength sensitivity and rapid response. Three telescopes are co-aligned.
The Burst Alert Telescope (BAT) \citep{Barthelmy} is a broad field of view (FOV)
coded-aperture instrument with a CdZnTe detector plane, designed to search for transient
events such as GRBs. Approximately 100 GRBs are discovered by BAT
each year. In response to a BAT trigger,
the spacecraft performs an autonomous slew to point the two narrow field of view instruments.
The Ultra-Violet and Optical Telescope (UVOT) \citep{Roming} performs follow-up
observations of GRBs in the 
170 to 600 nm band, with a
0.28$\deg\times$0.28$\deg$ field of view.
The X-Ray Telescope (XRT) \citep{Burrows} has an effective area peaking at 
110 cm$^2$ (at 1.5 keV),
a 0.4$\deg\times$0.4$\deg$ FOV, and an energy bandpass
of 0.3--10 keV. While spacecraft slewing can be initiated autonomously, achieving
repointing within approximately 2 minutes, the process can also be initiated from the ground.
The Swift Mission Operations Center (MOC), located near the University Park campus of
the Pennsylvania State University, receives over 1000 requests for ToOs from the astronomical
community each year. Response time to a ToO depends on
scientific priority and urgency. Due to communication limitations and human-in-the-loop
commanding, observation of the highest priority ToOs is typically achieved within 4 hours,
although frequently the response time is less than 1 hour. In this way, a ground-based
observatory such as LIGO-Virgo can provide the trigger for a transient event, with Swift
providing prompt follow-up observations in the X-ray, UV, and optical bands.

\section{Selection of gravitational-wave transients}

The LIGO-Virgo EM follow-up program took place during two observing periods
spanning a bit over two months
from 2009 December 17 to 2010 January 8 and from 2010 September 2 to 2010 October 20.
The details of the implementation and testing of the
end-to-end search on the GW end have been described elsewhere \citep{s6emmethods}.
Here we will summarize the key features and expand on aspects that were specific
to the joint program with Swift.

\subsection{Event selection}

GW candidate events were selected for follow-up based primarily
on their False Alarm Rate (FAR), the rate at which an event of equal or
greater significance is expected to occur in the absence of a true signal.
For this run, in order to trigger a Swift follow-up, we set a nominal
threshold on the candidate events' FAR at no more than one event per 35 
days of triple coincident
running (i.e., the GW instrument configuration when all three detectors were
acquiring sensitive data).

To avoid trying to image unusually poorly localized events, an additional
constraint was placed on candidates that
20\% or more of the {\it{a posteriori}} (see Section~\ref{sec:weighting})
weighted probability in the event skymap must
be covered by up to five 0.4$\deg\times$0.4$\deg$ tiles selected for
follow-up as Swift fields.
The number of fields picked for the Swift follow-up program reflected a
compromise
between the need to cover as large an area in the sky as possible and the
requirement to be 
minimally disruptive of Swift's other science targets. If five fields were
targeted by Swift
for each LIGO-Virgo candidate, the correct location of GW events
of high significance could be imaged with a probability greater than 50\% \citep{s6emmethods}.
Additionally, manual and automated checks were performed on data
quality in each interferometer before sending any alerts, eliminating events
which had an obvious problem associated with them.  For events which passed
all data selection criteria, observation requests were sent to Swift through 
web-based ToO submission\footnote{https://www.swift.psu.edu/secure/toop/too\_request.htm}.

During the course of the EM follow-up program, GW candidate transient events
were selected by transient-finding algorithms for un-modeled bursts as
well as compact binary star coalescences \citep{s6emmethods}.
Two such events met all criteria and were submitted for
follow-up with Swift. These were
identified by one of the generic transient-finding algorithms, called 
coherent WaveBurst \citep{coherentwaveburst}.
This algorithm uses wavelet decomposition to search for GWs
without relying on specific waveform models.
The ``January''  event occurred at 8:46 UTC on 2010 January 7. This was an 
event close to the end of the first LIGO-Virgo observation period, during 
which we lowered the nominal thresholds to 1 event per day for the FAR 
and 10\% for the weighted probability
of the event skymap. The thresholds were adjusted as the first LIGO-Virgo observation period
was approaching an end in order
to exercise the follow-up process at least once. 
The ``September'' event, which occurred at
6:42 UTC on 2010 September 16, passed all nominal criteria for follow-ups.
This event was later (in March 2011) revealed to be a ``blind injection''
artificially inserted into
the interferometers as a test of our detection
procedures \citep{s6cbc,bigdog,s6burst}. While
both of these events were ultimately tests of the system rather than plausible
candidates, they demonstrate the viability of performing rapid follow-ups of
potential GW signals using Swift.

\subsection{Position errors and tiling\label{sec:weighting}}

The typical uncertainty in sky location of a GW signal is large
(typically tens of square degrees) relative to the FOV of Swift's XRT and 
UVOT instruments.  This may be in part addressed by imposing the requirement 
for the reconstructed sky location to overlap with nearby galaxies.
We have used information from the Gravitational Wave Galaxy Catalog
(GWGC) \citep{GWGCcatalog} 
in order to fold into the tiling algorithm location, extent and
blue luminosity
of known galaxies within a distance less than 50 Mpc, since the GW 
interferometer network would not be likely to detect neutron star binary
coalescences beyond this distance.  
In this way, locations with a known galaxy from this catalog 
were given greater weight than those without galaxies present.

For the earlier ``winter'' run in which the January event
occurred, skymap tiles with galaxies or globular clusters at a distance less
than 50 Mpc  had their
estimated probability increased by a factor of 3, whereas for the later
``autumn'' run
the estimated relative probability of each tile in the skymap was calculated 
according to:

\begin{equation}
\label{skymap_computation}
P \propto \displaystyle\sum\limits_{i} \frac{M_i L}{D_i}
\end{equation}

\noindent where $M_i$ is the blue light luminosity of a 
galaxy (a proxy for 
star formation rate~\citep{Phinney1991,kop2008,cbcrates}),
$L$ is the likelihood from only the GW skymap
and $D_i$ is the distance of the galaxy from Earth \citep{Nuttall}. 
The index $i$ sums over each galaxy associated with the skymap tile.  Only $\sim$8\% of tiles in a
typical skymap were associated with one or more galaxies, while P was set to zero for tiles with
no associated galaxy \citep{s6emmethods}.  After this summation was performed, the resulting likelihoods
were renormalized to a probability of unity over the entire skymap.
In this way we constructed the {\it{a posteriori}} probability skymaps that
we used for prioritizing the Swift observations.

\begin{figure}
\hbox{\includegraphics[width=0.5\textwidth]{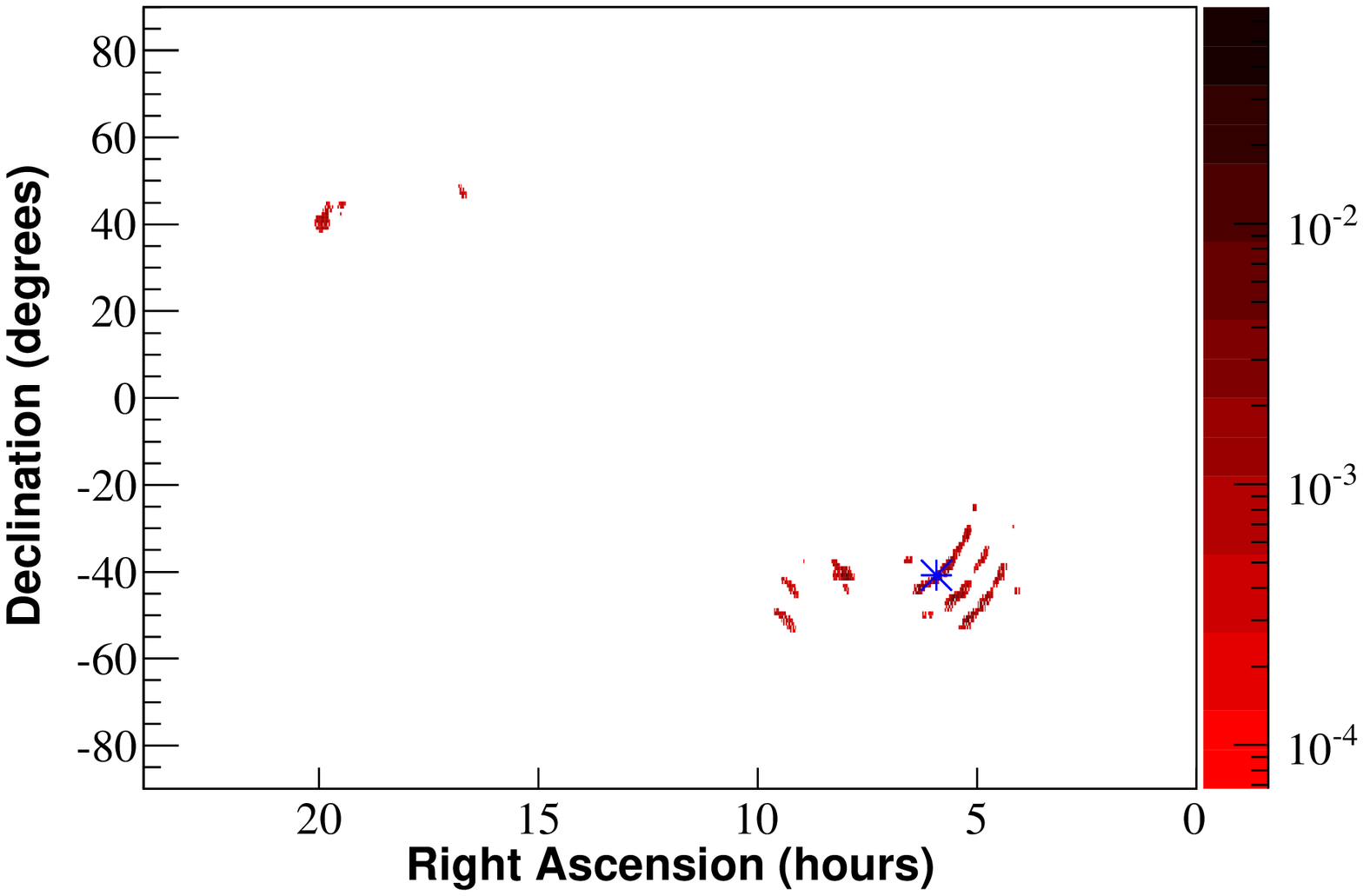}}
\hbox{\includegraphics[width=0.5\textwidth]{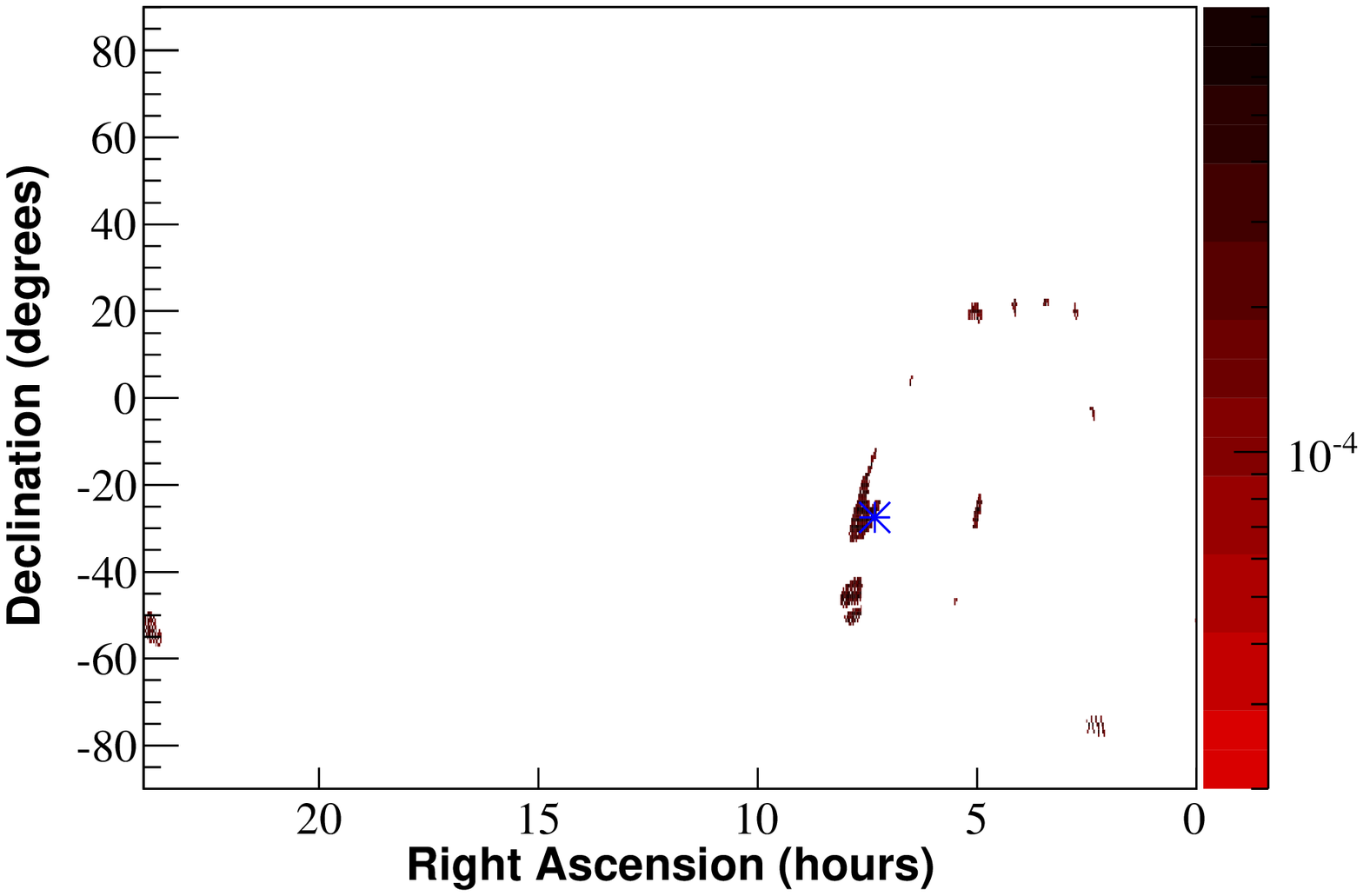}}
\caption{Probability skymaps of the January event (top) and September event (bottom) as determined
by the coherent WaveBurst algorithm before galaxy weighting. The asterisks mark the approximate locations of the locations selected for follow-up with Swift.
  Color scale shows fraction of total (raw) probability contained within that 
particular skymap tile.}
\label{fig:skymaps}
\end{figure}

For both events selected for follow-up, several tiles were chosen and passed
on for imaging as Swift fields according to the procedures outlined above.
The probability skymaps produced by coherent WaveBurst without galaxy weighting and containing the top 1000 tiles are shown in Figure \ref{fig:skymaps}. 
The maps show large spreads in probability on the sky, the first due to their
relatively low significance, the
second due to worse sensitivity in Virgo relative to the LIGO interferometers.
Simulations show that, despite the relatively small size of Swift fields with 
respect to the extent of our probability skymaps, sources originating in or
close to nearby galaxies can be correctly localized with reasonable success 
rates (see \citet{s6emmethods} and discussion in Section 5).
The regions containing the fields imaged by Swift are shown with an asterisk 
on each skymap, with precise coordinates of each field given in 
Tables~\ref{tab:JanObs}--\ref{tab:SepObs}.

\section{Observations with Swift}

\begin{figure}
\hbox{\hfill\includegraphics[width=0.5\textwidth]{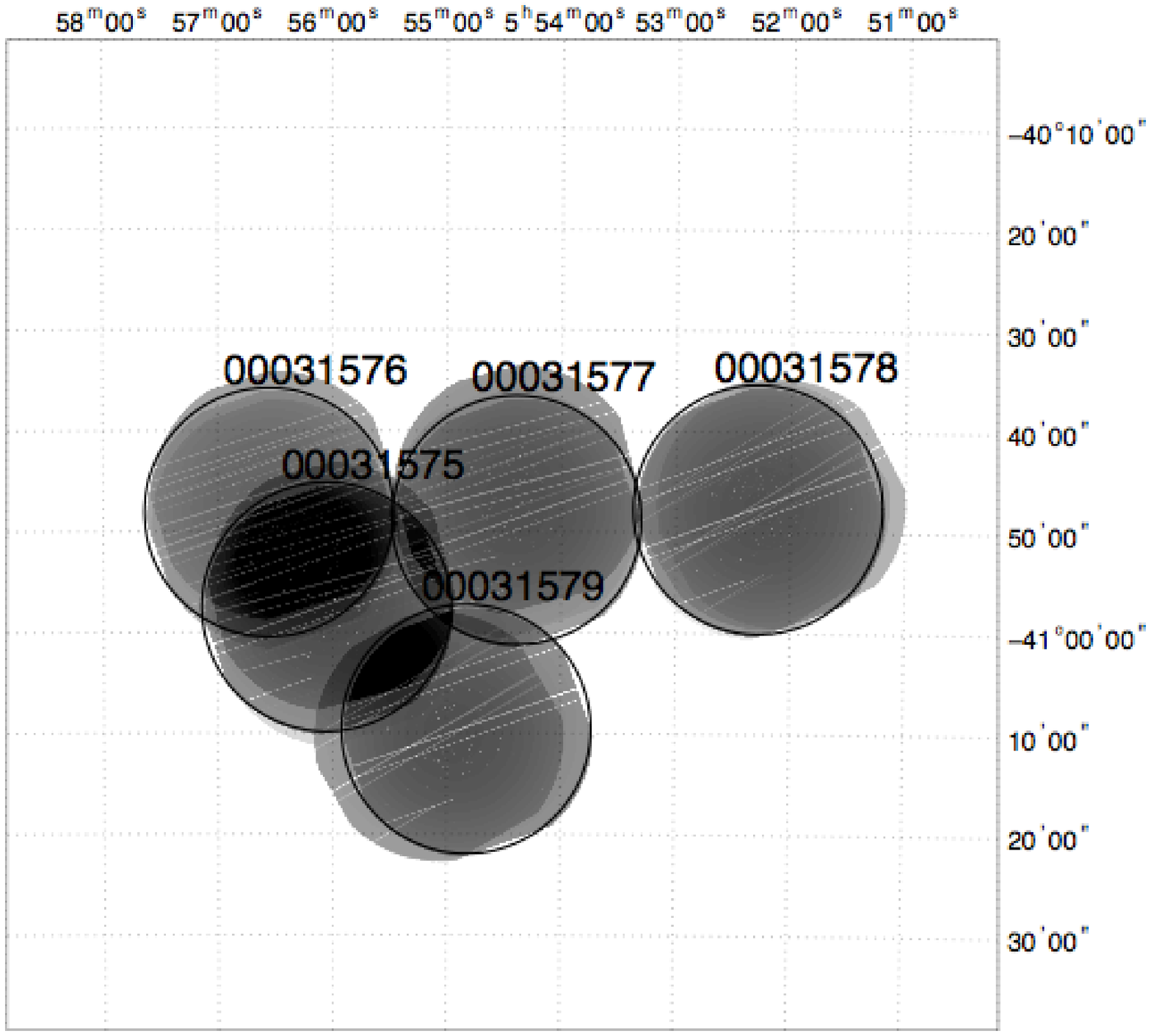}}
\hbox{\hfill\includegraphics[width=0.5\textwidth]{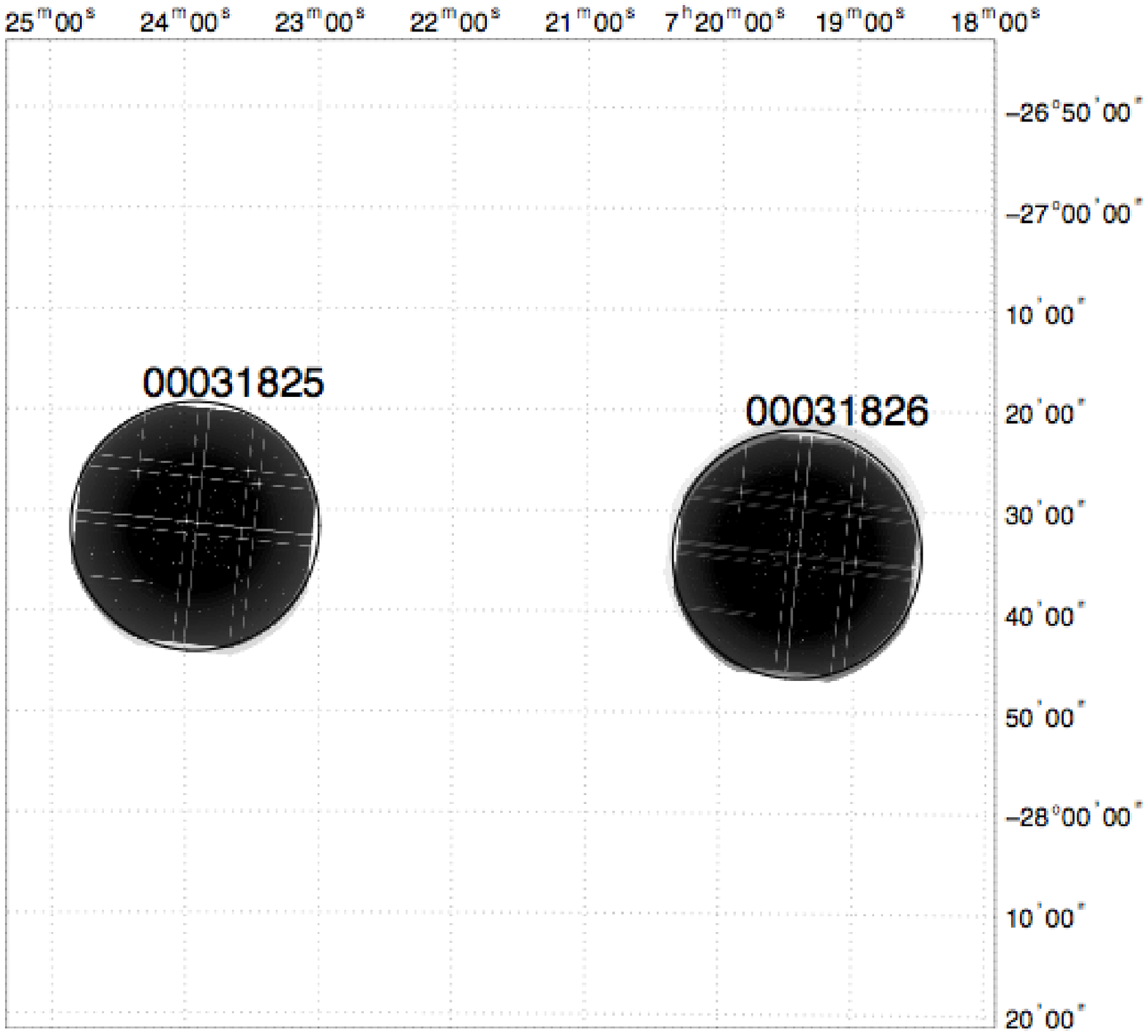}}
\caption{Swift-XRT image of the vicinity of the January event (top)
and September event (bottom). The large circles show the fields of view
of the various pointings. The grayscale denotes relative length of exposure
at each location with darker colors indicating longer ones.
The horizontal axis denotes right ascension (RA)
in hours, minutes and seconds and the vertical axis denotes declination (dec) 
in degrees.}
\label{fig:fields}
\end{figure}

The ToO requests were submitted manually via the web within 2 hours
from the collection of GW data.
The initial two requested pointings for each event were submitted as ``high priority'' 
(but not the highest);
this reflected a compromise between capturing the early light and being
minimally disruptive to Swift operations.
Swift observations with the XRT and UVOT instruments
of the specified targets were performed in
the following $\sim 12$ hours.
Rapid follow-up could significantly increase the likelihood of a detection, as
several types of potential GW/EM sources (e.g.\ short GRBs) have afterglows that fade below 
detectability on a timescale of $\sim$hours.
The highest probability region of the January event skymap was observed 
by Swift in 5 overlapping fields (Figure \ref{fig:fields}). Each of these were
observed twice that month. For the September event two disjoint Swift
fields were observed (Figure \ref{fig:fields}). Note that the fields shown are 
explicitly for the XRT analysis, whereas the UVOT has a slightly smaller FOV. 
Each September event field was observed on two days,
one in September 2010 and the other in December 2010. Details of the
observations are given in Tables~\ref{tab:JanObs}--~\ref{tab:SepObs}.

\subsection{X-ray results}

We analyzed the Swift-XRT data with custom scripts which use the
software described in \cite{EvansLC} and \cite{EvansCatalogue}.
For each field we
combined the observations and produced a single image and exposure map.
We then used the source detection and point spread function (PSF)
fitting code described in \cite{Goad}
and \cite{EvansCatalogue} to identify and localize sources
in the field. This method uses a sliding-cell detection algorithm, with
a fixed cell size of 21$\times$21 pixels
(49.6$^{\prime\prime}\times$49.6$^{\prime\prime}$, which encloses 93\% of the PSF). 
An initial run with a
detection significance threshold  of 3
(i.e.\ $S/\sigma_S$ is at least 3, where $S$ is the estimated number of
net source counts in the cell, and $\sigma_S$ is the uncertainty in this
value, determined using the background estimated from an annular box of
size 51 pixels)
revealed no
sources in any of the observations, however for faint sources this box
size is sub-optimal. We performed a second source search with
a reduced detection threshold of 1.5$\sigma$. We measured the mean
background level in
each field and, using a circular source region of radius 10 pixels
centered on the position of each of these ``reduced-threshold'' detections, we
applied the Bayesian test of \citet{Kraft}, only
accepting sources which this test determined to be detected with at
least 99.7\%\ confidence (i.e.\ 3$\sigma$ detections). While this
method makes us more sensitive to faint sources, it also increases
likelihood of getting false positives due to background inhomogeneities,
thus these detections should be treated with caution. A detection
system optimized for faint sources is under development. The positions
of the reduced-threshold detections are given in
Tables~\ref{tab:JanDet}--\ref{tab:SepDet}.

Each of these reduced-threshold detections had very few photons,
therefore it was not useful to perform a detailed spectral
analysis. We determined the mean count-rate of each detection using the
Bayesian method of \cite{Kraft} and then applied corrections for
PSF losses and instrumental effects, following the processes described
by \cite{EvansLC} and \cite{EvansCatalogue}. We used
PIMMS\footnote{http://heasarc.nasa.gov/Tools/w3pimms.html}
to determine a count rate-to-flux conversion factor, assuming an absorbed
power-law spectrum with a photon index of 1.7. For the January event
the assumed absorbing column was $4\times 10^{20}$ cm$^{-2}$, 
giving a 0.3--10 keV conversion factor of 4.2 $\times 10^{-11}$
erg cm$^{-2}$ ct$^{-1}$; for the September event the absorbing column
was $3\times 10^{21}$ cm$^{-2}$, giving a 0.3-10 keV conversion factor of
5.1$\times 10^{-11}$ erg cm$^{-2}$ ct$^{-1}$. A change of 0.2 in the assumed
index of the power-law spectrum would result in a $\sim$10\% change in these
conversion factors.  The absorbing columns were
taken as the Galactic values in the direction of the event, determined
from \cite{Kalberla} assuming the abundances of \cite{Anders}.

Given the sky area covered by the Swift observations (0.126 square
degrees per XRT field after instrumental corrections are applied) we
expect to find a number of serendipitous sources
in the XRT data. To quantify the likelihood of serendipitous source detection, 
we used the 2XMMi-DR3 catalogue \citep{XMMCat}, which is substantially
dominated by serendipitous X-ray sources found by the XMM-Newton observatory 
\citep{Jansen2001}.

We selected from this catalogue all
unique good sources (i.e.\ with a quality flag of 0; for the flag
definition see \citet{XMMCat}). For consistency with 2XMMi-DR3 we convert the
0.3--10 keV XRT count-rates into 0.2--12 keV fluxes, using an absorbed
power-law spectrum with a column of $N_{\rm H}=3\times10^{20}$ cm$^{-2}$
and a power-law photon index $\Gamma=1.7$, as used for
2XMM\footnote{http://xmmssc-www.star.le.ac.uk/Catalogue/2XMM/UserGuide\newline\_xmmcat.html\#TabECFs}.
For each XRT source we then counted the number of 2XMMi-DR3 sources with
values of the 0.2--12 keV flux---as measured with the
EPIC MOS-1 camera \citep{Turner2001} onboard XMM-Newton---at
least as bright as the source in question. This was then scaled by the ratio of the Swift-XRT
instantaneous field of view to the 2XMMi-DR3
unique sky coverage area (504 square degrees). This yielded, for each
Swift source, the number of serendipitous sources (N$_{s}$) of at least that brightness
expected in a single Swift field of view.  These values, along with the count
rates, are given in
Tables~\ref{tab:JanDet}--\ref{tab:SepDet}. For the January
event we expected a total of 7.5 serendipitous sources, compared to our
8 detections; for the September event we expected 17 serendipitous
sources, and had 12 detections. 
The number of serendipitous sources expected will not be a strong function
of Galactic latitude at these latitudes ($-27.6\deg$ for the January
event and $-6.1\deg$ for the September event) and flux levels;
Galactic source sky
densities are comparable to extragalactic values at a flux of around 
$5 \times 10^{-14}$ erg cm$^{-2}$ s$^{-1}$ only at much lower latitudes, 
see e.g.\ \citet{Motch}.

In preparing for the event collection and Swift follow-up, we
constructed a utility
catalog that would allow us to look up the X-ray history of
each field selected for observation in order to help determine if any
of the observed sources
coincide with known steady-state or variable sources. 
This X-ray source catalog, which we call
XGWC\footnote{http://aquarius.elte.hu/XGWC/index.html},
combines public data from the HEASARC Master X-ray
Catalog\footnote{http://heasarc.gsfc.nasa.gov/W3Browse/all/xray.html}
and the GWGC.
Upon examination of XGWC/HEASARC for known X-ray sources
within the observed fields
we found two X-ray sources (one of which was observed twice by Swift)
in the catalog which were within the fields for the January event.
Both were of unknown type.
These catalog sources are likely to be associated with the reduced-threshold
sources \#3 and \#4 reported in Table~\ref{tab:JanDet}.
The fields that were observed for the September event contained
only one XGWC/HEASARC source of unknown type.
This is likely to be associated with the reduced-threshold
source \#8 reported in Table~\ref{tab:SepDet}.

We also performed a variability test on the observed sources, since an EM 
counterpart to the LIGO-Virgo event may be expected to be fading. To do this
we produced a light curve of each source, creating one bin per Swift observation ID\footnote{For the September event we excluded the second
observation of each source, since this was very short.}, and determining the 
count-rate using
the Bayesian method \citep{Kraft}. The significance of any
variability was then found simply by dividing the difference in the two
bins by the errors in the two bins, added in quadrature. This is given
in Tables~\ref{tab:JanDet}--\ref{tab:SepDet}; two sources show moderate
(between 2 and 2.5$\sigma$) evidence for variability (January Source \#1 and September Source \#6).

In addition to the observations reported above, a further follow-up
observation was performed for the January trigger on 2010 Jan 30. This
observation, obs ID 00031592001, contained 5.4 ks of data, centered at
RA=88.73\deg, dec=-40.96\deg, i.e.\ overlapping heavily with field
00031577. Including the data from this field causes Sources \#1 and \#4 to
be registered as above-threshold detections.
For Source \#4, the count-rate of the final observation was almost
identical to the first; inclusion of this result changed the variability
significance from 0.75 to 0.04$\sigma$. For Source \#1, the addition of these
data alters the variability only slightly, from 2.05 to 1.98$\sigma$.

\begin{table*}
\caption{Details of the Swift-XRT follow-up observations of the January event.}
\label{tab:JanObs}
\begin{center}
\begin{tabular}{ccccc}
\hline
Swift Observation ID         &    Date Start         & Date End                & Exposure  & Pointing Direction \\
              &    (UT)               &  (UT)                   &   (ks)      & J2000\\
\hline
00031575001   & 2010 Jan 07 at 13:04   & 2011 Jan 07 at 14:29    &  1.5        & 89.07\deg, -40.96\deg\\
00031575002  &  2010 Jan 11 at 19:11   &  2010 Jan 11 at 19:38    & 1.6         & 89.00\deg, -40.92\deg \\
\\
00031576001  &  2010 Jan 07 at 14:33   &  2010 Jan 07 at 15:55    & 1.7         & 89.14\deg, -40.80\deg \\
00031576002  &  2010 Jan 11 at 15:59   &  2010 Jan 11 at 16:18    & 1.1         & 89.13\deg, -40.77\deg \\
\\
00031577001  &  2010 Jan 07 at 15:57   &  2010 Jan 07 at 17:20    & 1.8         & 88.60\deg, -40.81\deg \\
00031577002  &  2010 Jan 11 at 17:35   &  2010 Jan 11 at 18:02    & 1.6         & 88.60\deg, -40.76\deg \\
\\
00031578001  &  2010 Jan 07 at 17:23   &  2010 Jan 07 at 18:49    & 2.0         & 88.07\deg, -40.79\deg \\
00031578002  &  2010 Jan 11 at 13:18   &  2010 Jan 11 at 15:01    & 1.4         & 88.06\deg, -40.78\deg \\
\\
00031579001  &  2010 Jan 07 at 20:26   &  2010 Jan 07 at 20:59    & 2.0         & 88.71\deg, -41.16\deg \\
00031579002  &  2010 Jan 11 at 18:03   &  2010 Jan 11 at 19:51    & 1.4         & 88.77\deg, -41.17\deg \\
\\
\hline
\end{tabular}
\end{center}
\end{table*}

\begin{table*}
\caption{Details of the Swift-XRT follow-up observations of the September event.}
\label{tab:SepObs}
\begin{center}
\begin{tabular}{ccccc}
\hline
Swift Observation  &    Date Start         & Date End                & Exposure  & Pointing Direction \\
              &    (UT)               &  (UT)                   &   (ks)      & J2000\\
\hline
00031825001   & 2010 Sep 16 at 18:10  & 2010 Sep 16 at 18:35    &  1.5        & 110.98\deg, -27.53\deg \\
00031825002   & 2010 Dec 30 at 03:17  & 2010 Dec 30 at 03:19    &  0.09       & 110.98\deg, -27.54\deg \\
00031825003   & 2010 Dec 30 at 00:06  & 2010 Dec 30 at 03:34    &  2.3        & 110.98\deg, -27.53\deg \\
\\
00031826001   & 2010 Sep 16 at 19 36  & 2010 Sep 16 at 20:09    &  2.0        & 109.86\deg, -27.57\deg \\
00031826002   & 2010 Dec 29 at 03:37  & 2010 Dec 29 at 04:50    &  0.14       & 109.85\deg, -27.54\deg \\
00031826003   & 2010 Dec 29 at 03:38  & 2010 Dec 29 at 05:09    &  1.9        & 109.86\deg, -27.58\deg \\
\hline
\end{tabular}
\end{center}
\end{table*}

\begin{table*}
\caption{The reduced-threshold detections in the X-ray data for the January event. }
\label{tab:JanDet}
\begin{center}
\begin{tabular}{ccccccc}
\hline
Source \#     &    Right ascension (RA)        & Declination (dec)                                               & Position Error &  Count Rate              & N$_{\rm s}$~$^1$ & Variability~$^2$     \\
              &    (J2000)   &  (J2000)                                          & ($^{\prime\prime}$ 90\% conf.) & (0.3--10 keV, ks$^{-1}$) &               & significance ($\sigma$)\\
\hline
1         & 05h 55m 1.00s    &    -40\deg 58$^{\prime}$ 00.8$^{\prime\prime}$    & 4.5  & 5.9$^{+1.5}_{-1.2}$ &     0.9            & 2.05  \\
2         & 05h 57m 4.80s    &    -40\deg 54$^{\prime}$ 45.4$^{\prime\prime}$    & 4.3  & 5.9$^{+2.1}_{-1.6}$ &     0.9            & 0.26  \\
3         & 05h 54m 12.72s   &    -40\deg 44$^{\prime}$ 05.8$^{\prime\prime}$    & 4.3  & 4.6$^{+1.5}_{-1.2}$ &     1.3            & 0.45  \\
4         & 05h 54m 59.29s   &    -40\deg 54$^{\prime}$ 19.6$^{\prime\prime}$    & 4.5  & 3.2$^{+1.3}_{-1.0}$ &     2.4            & 0.75  \\
5         & 05h 51m 57.66s   &    -40\deg 46$^{\prime}$ 10.9$^{\prime\prime}$    & 5.6  & 2.8$^{+1.8}_{-1.1}$ &     2.9            & 1.10  \\
6         & 05h 51m 41.12s   &    -40\deg 44$^{\prime}$ 46.4$^{\prime\prime}$    & 5.5  & 1.4$^{+1.1}_{-0.7}$ &     7.5            & 0.74  \\
7         & 05h 52m 6.29s    &    -40\deg 59$^{\prime}$ 14.3$^{\prime\prime}$    & 6.5  & 2.3$^{+1.2}_{-0.8}$ &     3.9            & 0.91  \\
8         & 05h 52m 55.88s   &    -40\deg 46$^{\prime}$ 14.9$^{\prime\prime}$    & 5.2  & 2.9$^{+1.7}_{-1.2}$ &     2.8            & 2.00  \\
\hline
\end{tabular}
\tablenotetext{1}{N$_{\rm s}$ is the number of 2XMMi-DR3 sources which are at least
as bright as the Swift source, which are expected in a single Swift field. See text for details.}
\tablenotetext{2}{The significance of the difference in count-rate between the difference epoch observations.}
\end{center}
\end{table*}

\begin{table*}
\caption{The reduced-threshold detections in the X-ray data for the September event.}
\label{tab:SepDet}
\begin{center}
\begin{tabular}{ccccccc}
\hline
Source \#     &    Right ascension (RA)        & Declination (dec)                                               & Position Error &  Count Rate              & N$_{\rm s}$ & Variability  \\
              &    (J2000)   &  (J2000)                                          & ($^{\prime\prime}$ 90\% conf.) & (0.3--10 keV, ks$^{-1}$) &               & significance ($\sigma$)\\ 
\hline
1         & 07h 23m 22.99s   &    -27\deg 26$^{\prime}$ 10.1$^{\prime\prime}$    & 4.4 & 2.8$^{+0.9}_{-0.7}$ &    2.9             & 1.47  \\
2         & 07h 23m 22.34s   &    -27\deg 33$^{\prime}$ 09.5$^{\prime\prime}$    & 4.4 & 2.3$^{+1.1}_{-0.7}$ &    3.9             & 1.09  \\
3         & 07h 23m 34.43s   &    -27\deg 23$^{\prime}$ 32.4$^{\prime\prime}$    & 5.4 & 2.4$^{+1.1}_{-0.8}$ &    3.7             & 1.47  \\
4         & 07h 24m 34.95s   &    -27\deg 31$^{\prime}$ 31.1$^{\prime\prime}$    & 6.1 & 1.8$^{+1.2}_{-0.7}$ &    5.5             & 1.01  \\
5         & 07h 23m 53.50s   &    -27\deg 23$^{\prime}$ 06.5$^{\prime\prime}$    & 4.4 & 0.6$^{+0.3}_{-0.2}$ &    17              & 1.30  \\
6         & 07h 24m 27.89s   &    -27\deg 35$^{\prime}$ 40.8$^{\prime\prime}$    & 6.5 & 2.3$^{+1.1}_{-0.7}$ &    3.9             & 2.48  \\
7         & 07h 23m 54.14s   &    -27\deg 42$^{\prime}$ 29.5$^{\prime\prime}$    & 6.4 & 2.2$^{+1.0}_{-0.7}$ &    4.2             & 1.20  \\
8         & 07h 19m 30.22s   &    -27\deg 45$^{\prime}$ 42.5$^{\prime\prime}$    & 4.1 & 8.8$^{+3.4}_{-2.4}$ &    0.5             & 0.44  \\
9         & 07h 19m 37.14s   &    -27\deg 33$^{\prime}$ 12.0$^{\prime\prime}$    & 5.2 & 2.4$^{+1.1}_{-0.8}$ &    3.7             & 0.60  \\
10         & 07h 19m 25.72s   &    -27\deg 31$^{\prime}$ 37.0$^{\prime\prime}$    & 5.8 & 0.9$^{+0.6}_{-0.3}$ &    12              & 0.36  \\
11         & 07h 19m 18.04s   &    -27\deg 25$^{\prime}$ 15.4$^{\prime\prime}$    & 5.0 & 1.7$^{+0.9}_{-0.6}$ &    5.9             & 0.97  \\
12         & 07h 19m 41.92s   &    -27\deg 39$^{\prime}$ 58.1$^{\prime\prime}$    & 5.0 & 1.6$^{+1.2}_{-0.7}$ &    6.4             & 1.02  \\

\hline
\end{tabular}
\end{center}
\end{table*}

\subsection{Optical and UV results}

\begin{table*}
\begin{center}
\caption{UVOT photometry for XRT detections in January event\label{t:membersJ}}
\begin{tabular}{ccccccc}
\hline
XRT Source No. & 00031576001 & 00031577001 & 00031577002 & 00031578001 & 00031578002 & 00031592001 \\
  & white & white & white & u & u & uvm2 \\
\hline
1  & ...     & ...       & ...  & ...     & ... &  19.83$\pm$0.09 \\
2  &  20.09$\pm$0.06 & ... & ...  & ...    & ... & ...\\
3  & ...     &  17.70$\pm$0.04 & 17.78$\pm$0.04 & ...    & ... & ...\\
4  & ...     &  19.50$\pm$0.06 &  ...   & ...    & ...   &  18.36$\pm$0.04\\
5  & ...     &  ...      & ...  & $>$21.53       & $>$21.33 & ...\\
6  & ...     &  ...      & ...  & $>$21.55       & $>$21.27 & ...\\
7  & ...     &  ...      & ...  & ...   & ... & ...\\
8  & ...     &  ...      & ...  & $>$21.52       & $>$21.40 & ...\\
\hline
\end{tabular}
\tablecomments{Column headings designate UVOT fields and filters used and table entries are observed magnitudes.  No XRT sources fell within the UVOT
field of view for observations 00031575001, 00031575002, 00031576002, 00031579001 or 00031579002.}
\end{center}
\end{table*}

\begin{table*}
\begin{center}
\caption{UVOT photometry for XRT detections in September event\label{t:membersS}}
\begin{tabular}{ccccc}
\hline
XRT Source No. & 00031825001 & 00031825003 & 00031826001 & 00031826003 \\
 & u & u & white & white \\
\hline
1   & ...      & ...       & ...      & ...       \\  
2   & $>$20.73       & $>$21.21        & ...      & ...       \\ 
3   & 14.11$\pm$0.02 & 14.11$\pm 0.02$ & ...      & ...       \\ 
4   & $>$20.76       & $>$21.22        & ...      & ...       \\ 
5   & 19.76$\pm$0.17 & 20.67$\pm 0.23$ & ...      & ...       \\ 
6   & $>$20.79       & 21.10$\pm 0.33$ & ...      & ...       \\ 
7   & ...      & ...       & ...      & ...       \\ 
8   & ...      & ...       & ...      & ...       \\
9   & ...      & ...       & 20.75$\pm$0.36     & 20.32$\pm$0.12  \\ 
10  & ...      & ...       & 18.76$\pm$0.07 & 18.90$\pm$0.04  \\ 
11  & ...      & ...       & $>$20.79       & ...       \\ 
12  & ...      & ...       & $>$20.84       & $>$21.65        \\ 
\hline
\end{tabular}
\tablecomments{Column headings designate UVOT fields and filters used and table entries are observed magnitudes.}
\end{center}
\end{table*}

\begin{table*}
\begin{center}
\caption{Potential Variable Stars in UVOT Data\label{t:UVOTS}}
\begin{tabular}{ccccccc}
\hline
Variable Source No.& Right ascension (RA)& Declination (dec)& Filter&
Mag& $\sigma_M$& Variability\\
 & (J2000)& (J2000)& & & & \\
\hline
     V1   & 05:56:22.06 & -40:58:38.3 & white & 19.77 &  0.018 &   8.03\\
     V2   & 05:56:15.71 & -40:56:06.8 & white & 20.51 &  0.026 &   9.73\\
     V3   & 05:56:04.51 & -40:51:50.3 & white & 20.50 &  0.020 &   7.14\\
     V4   & 05:55:54.64 & -40:59:55.3 & white & 19.29 &  0.017 &   6.35\\
     V5   & 05:55:43.65 & -40:55:51.9 & white & 21.04 &  0.029 &   6.88\\
     V6   & 05:55:50.23 & -40:45:40.8 & white & 19.38 &  0.063 &  12.10\\
     V7   & 05:55:05.25 & -41:04:07.9 & white & 16.53 &  0.008 &  14.93\\
     V8   & 07:24:20.67 & -27:33:36.5 & u     & 17.26 &  0.013 &   9.67\\
     V9   & 07:23:50.86 & -27:30:24.8 & u     & 17.82 &  0.024 &   8.74\\
    V10   & 07:23:39.28 & -27:25:38.5 & u     & 19.42 &  0.030 &   7.29\\
    V11   & 07:19:08.99 & -27:37:35.0 & white & 16.27 &  0.044 &  14.50\\
\hline
\end{tabular}
\end{center}
\end{table*}

The LIGO-Virgo target fields were observed by the Swift UVOT with the clear white
filter ($160-800$~nm)
or, if ground analysis indicated that the field
contained a star too bright for observation in white, the broadband
$u$ filter (centered at $346.5$~nm with full width at half maximum
of $78.5$~nm) or narrow NUV $uvm2$
filter (centered at $224.6$~nm with full width at half maximum of $49.8$~nm).

We first attempted to identify any UVOT counterparts to the X-ray
reduced-threshold sources identified in
the XRT images, photometering any source within 5$^{\prime\prime}$ of the XRT position
using the standard
UVOTSOURCE code and calibrations from \citet{P08} and \citet{Breeveld}.
The results are listed in Tables \ref{t:membersJ}--\ref{t:membersS}.
For the January event, XRT reduced-threshold
detection 7 fell off the smaller UVOT field
and was not measured. UVOT counterpart sources to XRT reduced-threshold 
detections 1, 2, 3 and 4 were found in at least one epoch.  However, they had 
corresponding sources in the Digital Sky Survey (DSS) and showed no photometric
variation beyond the 1-2$\sigma$ level.  No counterparts to XRT
reduced-threshold detections 5, 6 and 8 were found in UVOT.

For the September event, XRT reduced-threshold detections 1, 7 and 8 fell 
outside of the UVOT field.  No counterparts to XRT reduced-threshold
detections 2, 4, 11 and 12 
were found by UVOT. Reduced-threshold detections 3 and 10 were also
found by UVOT and correspond to DSS sources.
XRT reduced-threshold detections 5, 6 and 9 correspond to very marginal
detections in a crowded field and are either very faint objects or
spurious detections.
The apparent dimming of the UVOT counterpart to source 5 is spurious as
this source was on the edge of one UVOT image and only a partial flux
could be measured.
In short, {\it no XRT reduced-threshold detection corresponds to either
an optical transient or a variable source} for either the January or
September event.

We next performed a blind search for variable targets within the
field. In this instance, we photometered the entire field using the DAOPHOT
\citep{Stetson87} PSF photometry program.  We photometered all data using a
quadratically variable PSF and found photometry and image subtraction to be
excellent, provided the PSF was based on stars not near the
coincidence-loss limit of the data.  Raw photometry was corrected for
coincidence loss, exposure time and large scale sensitivity. The number of
detections in each field ranged from 250 to 2800 depending on the Galactic
latitude of the field and the exposure time.  Of the nearly 6800 sources that
were detected in the UVOT data, approximately 5200 are well-measured
point-like sources that are not near the coincidence-loss limit off the
data.  Of these, 11 are not near chip edges or bright stars but show very
significant variability ($>7\sigma$) between the two epochs.
Their coordinates are shown in Table \ref{t:UVOTS}.
Such
significant variability is likely to reflect real phenomena, such as active
galaxies or variable stars, which are expected to be detected in any deep
field.  However, none of these variable sources corresponded to any X-ray
detections and none show variability beyond a few 0.1 magnitudes, which
would be consistent with a tentative classification as normal variable
stars or active galaxies.  Further monitoring would be needed to determine
their nature.

\subsection{Summary of EM findings}

The XRT analysis produced 20 reduced-threshold detections and the UVOT
analysis 
identified nearly  6800 sources in the follow-ups of the two events. 
However, all observations in X-ray, UV and optical bands were consistent with 
expectations for serendipitous sources and no single source displayed 
significant variability in the XRT or UVOT analyses.  

\section{Combined GW-EM results}
\label{combined_gw_em_results}

Information from the EM observations associated with a GW candidate event will have to be
combined with GW data in order to establish key quantities associated with a combined GW-EM
search for transients, like event significance or astrophysical reach.
In this section we present a formalism for combining results from such joint observations.
For the purpose of its validation we performed simulations using 
possible models of GW
and EM signals. In the process we verified the overall search procedure and 
estimated the increase in sensitivity of the search which resulted from the Swift follow-up observations. Although this is presented within the context of the LIGO-Virgo-Swift search, it can be straightforwardly extended to other joint searches as well.

In the way the search was conducted,
a typical joint candidate event is characterized by the strength of the
GW signal
expressed in terms of the coherent network
amplitude $\eta$ which is the detection statistic itself for coherent
WaveBurst. This quantity is generally proportional to the signal-to-noise ratio
and is described in detail in \cite{coherentwaveburst} and in \cite{s6burst}.
Joint candidate events are also characterized by
the measured X-ray flux, $S$,
and the sky location of the X-ray source, $\Omega \equiv$ [RA,dec].
In addition to these,
a skymap, i.e.\ the probability distribution for the location of a
potential GW source,
$p_{\mathrm{m}}(\Omega)$, is produced and used to select fields for imaging. We define the joint detection
statistic for an event as the logarithm of the joint likelihood ratio, sometimes also referred to as the Bayes factor, given by

\begin{equation}
\label{joint_lr}
\Lambda_{\mathrm{joint}}(\eta, S, \Omega) = \Lambda_{\mathrm{GW}}(\eta)\Lambda_{\mathrm{EM}}(S)\Lambda_{\mathrm{cor}}(\Omega).
\end{equation}             
In the above equation $\Lambda_{\mathrm{GW}} = p(\eta \given \signal)/p(\eta \given \noise)$ is the likelihood ratio for a GW candidate,
measuring its significance. $\Lambda_{\mathrm{EM}}(S) = p_0^{-1}(S)$ is the inverse of the
probability density of observing an accidental, serendipitous X-ray source which is not
correlated with the GW signal. The remaining term $\Lambda_{\mathrm{cor}}(\Omega) = p_{\mathrm{m}}(\Omega)$ is the probability
for a GW source to be in the location of the X-ray source, which measures positional
correlation between GW and EM signals. Detection statistics based on the likelihood ratio or the Bayes factor construction have been previously suggested and used in the context of searches for GW bursts \citep{Clark2007, Cannon2008, VelaGlitch} and GWs from compact binary coalescence \citep{LowMassCBCS5VSR1, Biswas2012}. In deriving Equation\ (\ref{joint_lr}) (see Appendix \ref{appendix:joint_lr} for details) we assumed that the dominant background in the EM sector is serendipitous X-ray sources that happen to be within the observed fields by chance. We neglected contributions to this from possible spurious sources - due to, e.g., instrumental artifacts - in the XRT analysis. In order to check if this assumption is justified we estimated from the 2XMMi-DR3 catalogue \citep{XMMCat} that one expects to find at least one serendipitous source with flux equal or greater than 5.4 $\times 10^{-13}$
erg cm$^{-2}$ s$^{-1}$ within five Swift fields.
Visual inspection of the XRT data would identify artifacts at or above
this flux level, and they would be excluded from the analysis. No such
artifacts were found in the fields analyzed in this paper.
We also assumed that inhomogeneities in the distribution of serendipitous sources over the sky are small.
As already mentioned earlier,
for every LIGO-Virgo candidate event Swift was nominally going to observe
five 0.4$\deg\times$0.4$\deg$ fields. The way the end-to-end search 
pipeline was constructed, only the most probable tiles according to the
sky-map $p_{\mathrm{m}}(\Omega)$ were observed as Swift fields.  
This resulted in a natural selection bias. As a result, the position correlation term, 
$\Lambda_{\mathrm{cor}}(\Omega)$, had a negligible effect
in separation of real events from background events. After verifying this through simulations
we drop it from the right hand side of Equation (\ref{joint_lr}), and the final expression for the detection statistic for the joint LIGO-Virgo and Swift search becomes

\begin{equation}
\label{detstat}
\rho_{\mathrm{joint}}(\eta,S) = \rho_{\mathrm{GW}}(\eta) + \rho_{\mathrm{EM}}(S)
\end{equation}       
where $\rho_{\mathrm{GW}}(\eta)$ and $\rho_{\mathrm{EM}}(S)$ are logarithms of $\Lambda_{\mathrm{GW}}(\eta)$
and $\Lambda_{\mathrm{EM}}(S)$ respectively.

We simulated the search by processing a population of model GW signals which
were paired with plausible X-ray fluxes.
The set of GW injections was the same with the one used 
previously \citep{s6emmethods} for methodological studies of
joint GW-EM observations.
These injections sampled the known galaxies within 50 Mpc according 
to the GWGC and were weighted to reflect each galaxy's blue light luminosity.
Their intrinsic strength (at the source) spanned {\it{ad hoc}} standard candle
values over 3 orders of magnitude.
For the purpose of this analysis and to approximate a more
realistic distribution of events that are 
relatively close to our detection threshold,
we imposed that they fall below detectability,
which implies $\eta \le 3.5$, at distances
outside the 50 Mpc range.
For the low latency
search in 2009-2010,  3.5 was the typical threshold value
for $\eta$. Translating this value into strain at the detector and
ultimately at the source depends mildly on the waveform morphology and
polarization state of the GW burst and rather strongly on the frequency
content of it \citep{s6burst}. In order
to set the scale,
at 50\% detection efficiency during the 2009-2010 run and for
GW bursts with energy content near 150Hz this value of $\eta$
corresponds to an isotropic
energy at the level of
$5.6\times 10^{-2}\solarmass$ emitted in GWs 
from a hypothetical source at
the Virgo cluster (16 Mpc) \citep{s6burst}.

All such simulated events were added to LIGO and Virgo data and
analyzed as in the actual search.
We used these simulated GW signals to compute the coherent network
amplitude, $\eta$, and its probability distribution, $p(\eta | signal)$,
which is needed for calculation of the joint statistic in
Equation\ (\ref{joint_lr}) and Equation\ (\ref{detstat}).
Models for X-ray counterparts were based
on GRB afterglows observed by Swift.
In order to set the scale of possible X-ray fluxes for counterparts,
we considered several short
hard GRBs and some bright and dim long GRBs~\citep{2009ApJ...703.1696Z}.
The host galaxies of all selected GRBs had
$z < 1$. We sampled the observed
X-ray afterglow light curves for the observed GRBs at different time lags
($\sim 10^4$ s, $10^5$ s, $10^6$ s) relative to the time of arrival
of the burst.
For a possible X-ray counterpart to GWs from a source at 50 Mpc
away this analysis gave us a wide range of flux values, $S_{50\mathrm{Mpc}}$,
from  $10^{-14}$ erg s$^{-1}$ cm$^{-2}$ to $10^{-8}$ erg s$^{-1}$ cm$^{-2}$.
In the absence of any other
guidance, we performed simulations for each order of magnitude in that range.
For a given $S_{50\mathrm{Mpc}}$ every GW signal was paired up with a
corresponding X-ray flux, 
which was scaled up (as it was positioned anywhere within the 50 Mpc range)
according to the distance to the source. 

Computation of the joint detection statistic, Equation\ (\ref{detstat}),
also requires estimates
of the background noise in GW detectors and flux distribution of serendipitous X-ray sources. Background noise in GW detectors is dominated by high amplitude instrumental artifacts.
It is typically estimated by time-shifting data from one detector with respect
to the other. In our simulations we used estimates for background noise from the coherent WaveBurst
search for GW bursts with LIGO and Virgo in their 2010 science run \citep{s6burst}.
In the EM sector, on the other hand, serendipitous X-ray
sources observed in coincidence with a GW signal are the main source of false alarms.
As in assessing the background in the XRT analysis we presented in Section 4.1,
we used also here the
2XMMi-DR3 catalogue \citep{XMMCat} of serendipitous sources for the
estimation of the
flux distribution for such sources. For both types of
backgrounds we fitted analytical models to the data and computed 
$p(\eta | \noise)$ and $p_0^{-1}(S)$ appearing in the
definition of the joint likelihood ratio in Equation\ (\ref{joint_lr}).
 
\begin{figure}
\hbox{\hfill\includegraphics[width=0.48\textwidth]{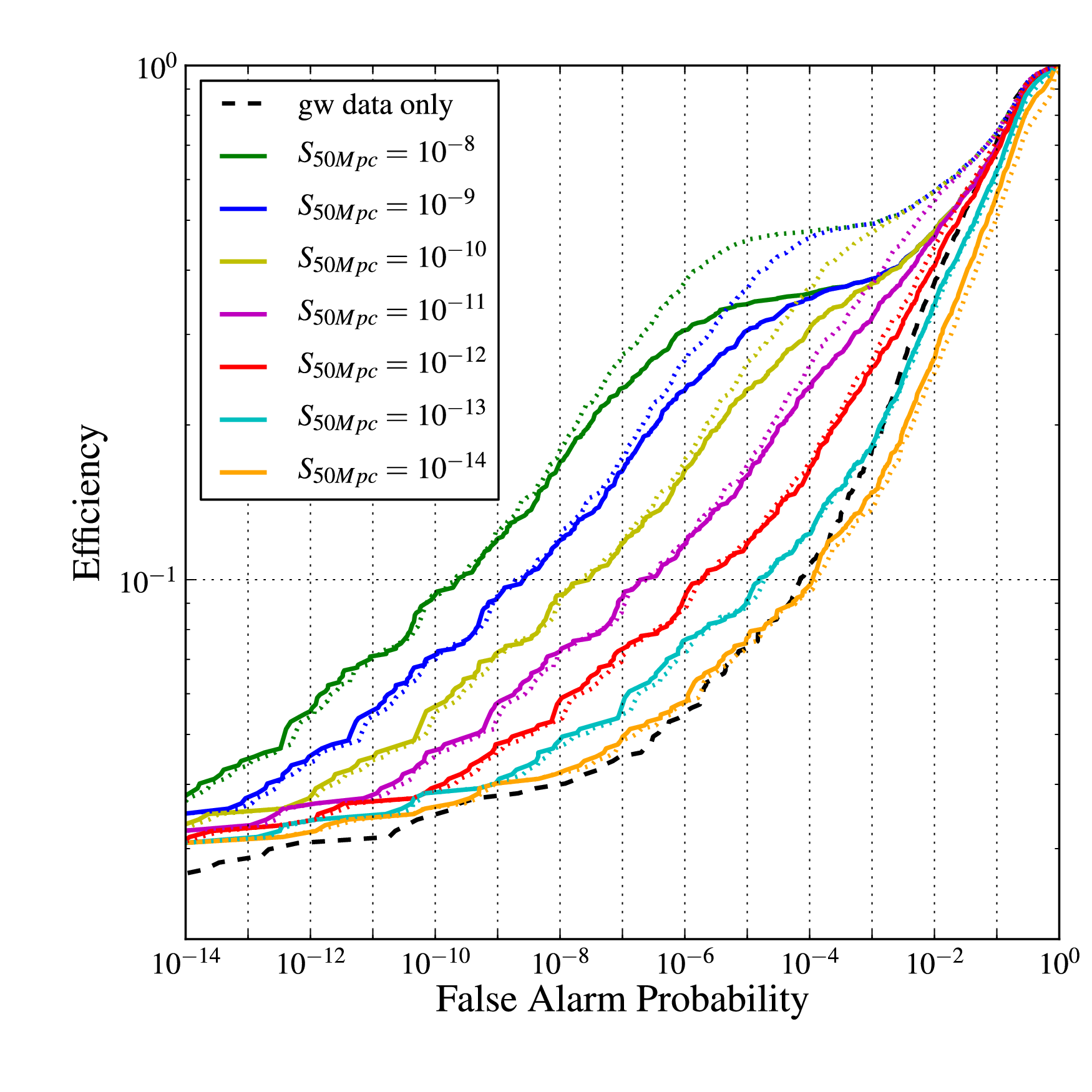}}
\caption{Efficiency as a function of false alarm probability for the joint LIGO-Virgo and Swift search. The solid (dotted) curves represent performance of the joint search with five (ten) fields observed by Swift for various values of the flux of an X-ray counterpart (in units of erg s$^{-1}$ cm$^{-2}$) at a distance of 50 Mpc, $S_{50\rm{Mpc}}$. The dashed line is the curve for the GW only search.}
\label{fig:roc}
\end{figure}

For each simulated GW signal complemented by an X-ray counterpart of a given
flux, the last step of the analysis involved the calculation of the
joint detection statistic $\rho_{\mathrm{joint}}$
as given by Equation\ (\ref{detstat}).
Using estimates for the GW and X-ray backgrounds, the false alarm
probability (FAP) of observing a background event with joint statistic
$\rho_{\mathrm{joint}}' \geq \rho_{\mathrm{joint}}$
in a month long search was computed.
This defined the significance of the observed event.
Figure \ref{fig:roc} shows the efficiency
in detecting these simulated GW-EM signals as a function of
FAP, $P_0$, in a joint
LIGO-Virgo-Swift search using five or ten Swift fields and for
a wide range of X-ray counterpart fluxes. The efficiency is defined as
the fraction of simulated signals
with FAP, $P_0' \leq P_0$.
For comparison, we also plot the efficiency curve for the GW
only search which does not use any (Swift) EM follow-up.
In the rare-event region
below FAP of $10^{-4}$ (i.e., corresponding to below $\sim4\sigma$
for the case of Gaussian statistics)
one can see how at fixed event significance the efficiency can improve by a
significant factor depending on the associated EM flux 
that is measured.
As expected, the gain increases with the brightness of the X-ray
counterpart. It reaches
a saturation point at which roughly one-third and one-half, respectively, 
of the signals
are detected in the searches with five and ten fields observed by Swift.
This is determined by the number of signals for which their true location
overlapped with the five (or ten) most
significant tiles of the sky-map.
Only these signals were observed by Swift.
The rest of the signals were ``missed'' in the follow-up and therefore
did not benefit from it. Observing more fields with Swift increases the 
chances of locating the X-ray counterpart, but at the same time has the 
negative effect of increasing chances of accidental detection of serendipitous
 X-ray sources (background).
Figure \ref{fig:roc} shows that the net gain is noteworthy if ten
instead of five fields were observed by Swift.
In general, an X-ray telescope with a wide field of view
would be optimal for the purpose of the joint search.
We should note though that for such a telescope the position correlation
term, $\Lambda_{\mathrm{cor}}(\Omega)$, in the joint likelihood ratio
Equation\ (\ref{joint_lr}) may become important and should be included
in the analysis. 

\section{Summary and Discussion}

During two periods in late 2009 and 2010 the LIGO-Virgo GW
interferometric
network sent out low latency candidate GW events to partner
observatories for rapid 
follow-ups in various EM bands.  Two such events were followed up as ToOs
by Swift.
One of the events followed up by Swift was ultimately revealed to be
a blind injection artificially inserted into the data as a test of the
system \citep{bigdog} and the other was a reduced threshold
test event \citep{s6cbc,s6burst}.
Prompt analysis of both the XRT and UVOT 
data obtained from the 7 total fields observed showed results consistent
with expectations for  serendipitous sources.
Given the lack of EM
candidates standing out above background, these 
particular observations do not increase our confidence in the
validity of the GW transients as established by the GW detectors alone.

Combining GW and EM astronomy will be pivotal in maximizing the science
in the advanced detector era of gravitational interferometers; it
may not only increase our confidence in the detection of
GWs but also complete our understanding of the astrophysics 
of the observed systems
\citep{Bloometal2009,Phinney2009,Stamatikos2009,Metzger2012}.
Our prototype observing program and end-to-end analysis has been the
first step in joint X-ray and GW observations. We demonstrated 
their feasibility and the considerable added value joint observations bring.
Improvements to these will continue to be made in the future on both the
EM and GW side.
The relatively narrow FOV of instruments
such as Swift with respect to the limited pointing resolution abilities
of GW interferometers 
makes identifying the position of the source on the sky non-trivial.
A possible fourth detector site in India, Japan or elsewhere and
continued refinements in
source localization algorithms are likely to
reduce the sky-position error area.
On the EM side, a more highly optimized faint source detection scheme for
XRT transients might yield improvement in EM sensitivity.
In late 2011 Swift implemented 
on-board software changes to allow automatic scheduling sequences of
partially overlapping 
XRT FOV exposures in response to ToO observation requests for targets with
position uncertainties larger than the FOV- this will assist the follow-up
of GW targets.
A significant role will also be played by prompt follow-up campaigns in 
the optical
band that may provide rapid sub-degree source localizations. Such
localizations may facilitate the subsequent follow-up with narrow FOV instruments
including Swift, thus significantly improving the chances of capturing the
X-ray signatures of GW sources.

In order to carry out multimessenger astrophysics with GWs, it will be
extremely important to have Swift 
and/or Swift-like satellites capable of rapid pointing,
multi-wavelength observations
and of as wide a field of view as possible operating concurrently with the
advanced GW detector network later in this decade.
Maximizing the 
science from GW astronomy will require sensitive partner instruments all
across the EM spectrum.
The successful completion of this
end-to-end program of EM follow-ups by Swift and other observatories during
the most recent science runs of the
LIGO-Virgo network provides confidence that such joint observations will be 
both technically feasible and scientifically valuable endeavors in the future.

\acknowledgments
The authors gratefully acknowledge the support of the United States
National Science Foundation for the construction and operation of the
LIGO Laboratory, the Science and Technology Facilities Council of the
United Kingdom, the Max-Planck-Society, and the State of
Niedersachsen/Germany for support of the construction and operation of
the GEO600 detector, and the Italian Istituto Nazionale di Fisica
Nucleare and the French Centre National de la Recherche Scientifique
for the construction and operation of the Virgo detector. The authors
also gratefully acknowledge the support of the research by these
agencies and by the Australian Research Council, 
the International Science Linkages program of the Commonwealth of Australia,
the Council of Scientific and Industrial Research of India, 
the Istituto Nazionale di Fisica Nucleare of Italy, 
the Spanish Ministerio de Econom\'ia y Competitividad,
the Conselleria d'Economia Hisenda i Innovaci\'o of the
Govern de les Illes Balears, the Foundation for Fundamental Research
on Matter supported by the Netherlands Organisation for Scientific Research, 
the Polish Ministry of Science and Higher Education, the FOCUS
Programme of Foundation for Polish Science,
the Royal Society, the Scottish Funding Council, the
Scottish Universities Physics Alliance, The National Aeronautics and
Space Administration, the Carnegie Trust, the Leverhulme Trust, the
David and Lucile Packard Foundation, the Research Corporation, and
the Alfred P. Sloan Foundation.
This work was also partially supported through a NASA grant/cooperative
agreement number NNX09AL61G to the Massachusetts Institute of Technology.
P. Evans and J.P. Osborne acknowledge financial support 
from the UK Space Agency.

\appendix

\section{Derivation of the joint likelihood ratio}
\label{appendix:joint_lr}

In this section we derive from first principles Equation\ (2) that we used in Section
5 in order to establish the detection statistic of the joint GW-EM search.
Using Bayes' theorem, the probability skymap for a potential GW source, $p_{\mathrm{m}}(\Omega)$, can be written as 
\begin{equation}
\label{skymap_definition}
p_{\mathrm{m}}(\Omega \given \eta, \signal) = \frac{p(\eta \given \Omega, \signal )p(\Omega \given \signal)}{p(\eta \given \signal)}
\end{equation}
where $p(\eta \given \Omega, \signal )$ and $p(\eta \given \signal)$ are the conditional probabilities to measure a GW signal with coherent network amplitude $\eta$ in the case of a source located at $\Omega$ and a source with unknown location, respectively; $p(\Omega \given \signal)$ is the prior probability distribution for source location, which in this search is determined by the distribution of galaxies in the GWGC catalog and distance weighting, see Equation\ (\ref{skymap_computation}).\\

\noindent The joint likelihood ratio is defined as 

\begin{equation}
\Lambda_{\mathrm{joint}}(\eta, S, \Omega) = \frac{p(\eta, S, \Omega \given \signal)}{p(\eta, S, \Omega \given \noise)} = \frac{\int p(\eta \given \Omega', \signal )p(S, \Omega \given S', \Omega', \signal)p(\Omega' \given \signal)p(S' \given \signal) \diff \Omega' \diff S'}{p(\eta \given \noise)p_0(S)}
\end{equation}
where $p_0(S)$ is the probability density of observing an accidental, serendipitous X-ray source which is not
correlated with the GW signal, and we introduce $p(S, \Omega \given S', \Omega', \signal)$, the probability distribution of flux, $S$, and X-ray source location, $\Omega$, as measured by Swift for a source whose true flux at the detector and location are $S'$ and $\Omega'$, respectively; $p(S'\given \signal)$ is the prior probability distribution for the flux of an X-ray counterpart to the GW signal. Integration in the numerator is performed over all possible values of flux and sky locations. Note that, although we assume the strength of the GW signal to be uncorrelated with the flux of the X-ray counterpart, we demand that both types of signals originate from the same sky position, $\Omega'$. This enforces correlation between measured locations of GW and its X-ray counterpart. \\
  
\noindent Using Equation\ (\ref{skymap_definition}) we can express the joint likelihood ratio in terms of the skymap, $p_{\mathrm{m}}(\Omega)$, and GW likelihood ratio, $\Lambda_{\mathrm{GW}}(\eta) = p(\eta \given \signal)/p(\eta \given \noise)$:   

\begin{equation}
\Lambda_{\mathrm{joint}}(\eta, S, \Omega) = \Lambda_{\mathrm{GW}}(\eta)\frac{\int p_{\mathrm{m}}(\Omega')p(S, \Omega \given S', \Omega', \signal)p(S' \given \signal) \diff \Omega' \diff S'}{p_0(S)}.
\end{equation}
 \noindent This expression can be simplified further by postulating $p(S, \Omega \given S', \Omega', \signal) = \delta(S-S') \delta(\Omega - \Omega')$ and $p(S' \given \signal)=1,$
 
 \begin{equation}
 \Lambda_{\mathrm{joint}}(\eta, S, \Omega) = \Lambda_{\mathrm{GW}}(\eta)\frac{p_{\mathrm{m}}(\Omega)}{p_0(S)}.
 \end{equation}
 Defining $\Lambda_{\mathrm{EM}}(S) = p_0^{-1}(S)$ and $\Lambda_{\mathrm{cor}}(\Omega) = p_{\mathrm{m}}(\Omega)$, we arrive at the form for the joint likelihood ratio given in Equation\ (\ref{joint_lr}). We stress that the particular  form of $\Lambda_{\mathrm{EM}}(S)$ and  $\Lambda_{\mathrm{cor}}(\Omega)$  is a consequence of the simplifying assumptions about Swift's ability to measure the flux and location of an X-ray source that are justified in Section \ref{combined_gw_em_results}. In general, these quantities are going to be nontrivial ratios of likelihoods estimating odds of an X-ray candidate source to be a counterpart to GW based on its brightness and location.

\end{document}